\documentclass[conference]{IEEEtran}

\usepackage{graphicx}
\DeclareGraphicsExtensions{.pdf,.jpeg,.png}
\usepackage{acronym}
\usepackage{array}
\usepackage{mdwmath}
\usepackage{mdwtab}
\usepackage{fixltx2e}
\usepackage{amsmath}
\usepackage{amssymb}
\usepackage{pgfplots}
\usepackage{tikz}
\usepackage{ifthen}
\usepackage{pdfcomment}
\usepackage[Gray,squaren,thinqspace,thinspace]{SIunits}
\usepackage{lipsum}
\usepackage{caption}
\usepackage{epstopdf}
\usepackage{amsthm}
\usepackage{subcaption}
\usepackage{todonotes}
\usepackage[indexonlyfirst,acronym,xindy,toc,nopostdot,nomain]{glossaries}
\usepackage{nicefrac}

\usepackage{color}
\definecolor{tud1d}{RGB}{36,53,114}
\definecolor{tud3d}{RGB}{0,113,94}
\definecolor{tud6d}{RGB}{174,142,0}
\definecolor{tud8d}{RGB}{169,73,19}
\definecolor{tud10d}{RGB}{115,32,84}

\newcommand{\mat}[1]{\ensuremath{\mathbf{#1}}}													
\newcommand{\stiff}{\ensuremath{\mat{K}}}														    
\newcommand{\parder}[2]{\ensuremath{\frac{\partial #1}{\partial #2}}}		

\newcommand{\norm}[2]{\ensuremath{\left\Vert #1 \right\Vert_{#2}}}      
\newcommand{\Ndof}{\ensuremath{N_\text{DoF}}}    			    					   	

\newcommand{\weight}{\ensuremath{\omega}}																

\newcommand{\Ort}{\ensuremath{\Omega_{\mathrm{rt}}}}
\newcommand{\Ost}{\ensuremath{\Omega_{\mathrm{st}}}}
\newcommand{\oOrt}{\ensuremath{\overline{\Omega}_{\mathrm{rt}}}}
\newcommand{\oOst}{\ensuremath{\overline{\Omega}_{\mathrm{st}}}}
\newcommand{\Gd}{\ensuremath{\Gamma_{\mathrm{d}}}}
\newcommand{\Gl}{\ensuremath{\Gamma_{\mathrm{l}}}}
\newcommand{\Gr}{\ensuremath{\Gamma_{\mathrm{r}}}}
\newcommand{\Gag}{\ensuremath{\Gamma_{\text{ag}}}}

\newcommand{\Azst}{\ensuremath{A_{z,\mathrm{st}}}}
\newcommand{\Azrt}{\ensuremath{A_{z,\mathrm{rt}}}}
\newacronym{fem}{FEM}{Finite Element Method}
\newacronym{iga}{IGA}{Isogeometric Analysis}
\newacronym{cad}{CAD}{Computer Aided Design}
\newacronym{nurbs}{NURBS}{Non-Uniform Rational B-splines}
\newacronym{pmsm}{PMSM}{permanent magnet synchronous machine}
\newacronym{pm}{PM}{permanent magnet}
\newacronym{thd}{THD}{total harmonic distortion}
\newacronym{emf}{EMF}{electromotive force}

\begin{document}
%
\title{Modelling of a Permanent Magnet Synchronous Machine Using Isogeometric Analysis} 


\author{\IEEEauthorblockN{Prithvi Bhat, Herbert De Gersem}
\IEEEauthorblockA{ Theorie Elektromagnetischer Felder\\
Technische Universit{\"a}t Darmstadt\\
Darmstadt, Germany}
\and
\IEEEauthorblockN{Zeger Bontinck, Sebastian Sch\"ops}
\IEEEauthorblockA{Theorie Elektromagnetischer Felder\\
Graduate School of CE\\
Technische Universit{\"a}t Darmstadt\\
Darmstadt, Germany}
\and
\IEEEauthorblockN{Jacopo Corno}
\IEEEauthorblockA{Theorie Elektromagnetischer Felder\\
Graduate School of CE\\
MOX Modeling and Scientific Computing \\ 
Politecnico di Milano, Italy}}

\IEEEtitleabstractindextext{%
\begin{abstract}
\gls*{iga} is used to simulate a permanent magnet synchronous machine. \gls*{iga} uses \gls*{nurbs} to parametrise the domain and to approximate the solution space, thus allowing for the exact description of the geometries even on the coarsest level of mesh refinement. Given the properties of the isogeometric basis functions, this choice guarantees a higher accuracy than the classical \gls*{fem}.

For dealing with the different stator and rotor topologies, the domain is split into two non-overlapping parts on which Maxwell's equations are solved independently in the context of a classical Dirichlet-to-Neumann domain decomposition scheme. The results show good agreement with the ones obtained by the classical finite element approach.
\end{abstract}

}

\maketitle

\IEEEdisplaynontitleabstractindextext

\IEEEpeerreviewmaketitle


\section{Introduction}
\label{sec:Intro}

Electric machines are usually modelled through the magnetoquasistatic approximation of Maxwell's equations discretised on the machine cross section, which requires the solution of a 2D Poisson problem. Its numerical solution commonly amounts to triangulating the domain and applying the \gls*{fem}. The acting electromotive forces and the torque can be determined by a post-processing procedure, e.g. invoking the Maxwell stress tensor method. A drawback of the standard approach is that a very fine mesh is needed to achieve an acceptable accuracy. Moreover, the air gap needs to be resolved properly and the solution is known to be extremely sensitive to the used discretisation~\cite{Howe_1992aa}. Furthermore, remeshing or reconnecting mesh parts in order to account for the machine rotation can introduce a spurious ripple on the solution for the torque~\cite{Tsukerman_1995aa}.

An exact parametrisation of the air gap is not possible within a classical \gls*{fem} framework since elements of any order rely on polynomial mappings which are unable to represent conic sections such as circles and ellipses. Furthermore, the regularity of \gls*{fem} solutions is typically limited by the $C^0$ continuity across the elements~\cite{Tarnhuvud_1988aa}.

\gls*{iga}~\cite{Hughes_2005aa} is able to overcome these issues. It chooses \gls*{cad} basis functions such as e.g. B-splines and \gls*{nurbs} for the approximation spaces and is thus able to represent \gls*{cad} geometries exactly even on the coarsest level of mesh refinement. A higher regularity at the mesh interfaces can also be obtained by applying $k$-refinement, which results in smoother basis functions compared to \gls*{fem} and increases the accuracy of the solution accordingly~\cite{Evans_2009aa}. Another benefit of IGA is that it provides an elegant way to do geometric optimisation~\cite{Pels_2015aa} and to handle uncertainties~\cite{Corno_2015aa}.

This work aims at using \gls*{iga} to model and simulate a \gls*{pmsm}. The stator and rotor models parts are inevitably resolved by multiple patches, which in turn necessitates the use of a domain decomposition approach across the air gap. We employ a classical iterative procedure based on a Dirichlet-to-Neumann map~\cite{Quarteroni_1999aa}.

In the following section, we introduce the 2D model electric machine model and the \gls*{iga} method used for its spatial discretisation. In section~\ref{sec:Coupling}, the iterative domain decomposition scheme is explained and, finally, in section~\ref{sec:Res}, we show the results for the simulation of a \gls*{pmsm}.

\section{Solving Maxwell's Equations for a \gls*{pmsm}}
\label{sec:maxwell}

\subsection{Magnetostatics}

\begin{figure}
\centering
\def\svgwidth{0.75\columnwidth}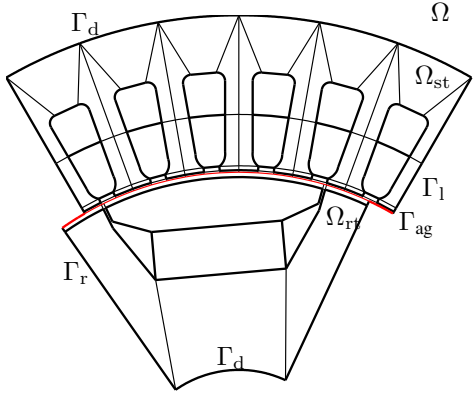
\caption{\label{fig:machine} One pole of the \gls*{pmsm}. In red, the interface $\Gag$ between the rotor and the stator is highlighted.}
\end{figure}

In early design steps, it is sufficient to model electric machines by the 2D magnetostatic approximation of the Maxwell's equations. Let $\Omega=\oOrt\cup\oOst$ (see Fig.~\ref{fig:machine}) depict the computational domain. One has to solve the following Poisson equation
\begin{equation}
\label{eq:poisson}
\left\{\begin{array}{rl}
-\nabla \cdot \left(\nu \nabla A_z\right)&=J_{\mathrm{src,}z}+J_{\mathrm{pm,}z},\\
A_z|_{\Gd} &=  0,\\
A_z|_{\Gl} &= -A_z|_{\Gr},
\end{array}\right.
\end{equation}
where $\nu=\nu(x,y)$ is the reluctivity, $A_z=A_z(x,y)$ is the $z$-component of the magnetic vector potential, $J_{\mathrm{src,} z}=J_{\mathrm{src,} z}(x,y)$ is the source current density in the coils and $(0,0,J_{\mathrm{pm},z})=(0,0,J_{\mathrm{pm},z}(x,y))=-\nabla\times\vec{H}_{\mathrm{pm}}$ is the current density according to the magnetisation $\vec{H}_{\mathrm{pm}}=\vec{H}_{\mathrm{pm}}(x,y)$ of the permanent magnets. On $\Gd$, a Dirichlet boundary condition is applied and on $\Gl$ and $\Gr$, an anti-periodic boundary condition holds.

Applying the loading method \cite{Rahman_1991aa} on the solution of \eqref{eq:poisson} gives us the spectrum of the \gls*{emf}. The first harmonic $E_1$ is the \gls*{emf} of the machines. The higher harmonics are used to calculate the \gls*{thd}, which is defined as
\begin{equation}
\label{eq:THD}
\text{THD}=\frac{\sqrt{\sum_{p=2}^\infty |E_p|^2}}{|E_1|},
\end{equation}
with $p$ the harmonic order and $E_p$ the \gls*{emf} of order $p$.

\subsection{Discretisation in the Isogeometric Framework}
\label{sec:IGA}
For the discretisation of $A_z$, a linear combination of basis functions $w_j=w_j(x,y)$ is used, i.e.,
\begin{equation}\label{eq:apprx}
A_z = \sum_{j=1}^{\Ndof} u_j w_j,
\end{equation}
where $u_j$ are the unknowns and $\Ndof$ depicts the number of unknowns. Using the Ritz-Galerkin approach, a system of equations is obtained, i.e.,
\begin{equation}\label{eq:SoE}
\stiff_\nu\mathbf{u}=\mathbf{j}_\mathrm{src}+\mathbf{j}_\mathrm{pm},
\end{equation}
where $\stiff_\nu$ has the entries
\begin{subequations}
\begin{equation}
k_{\nu,ij} = \int_\Omega \left(\nu\parder{w_i}{x}\parder{w_j}{x}+\nu\parder{w_i}{y}\parder{w_j}{y} \right)\;\text{d}\Omega,
\end{equation}
and the entries for the discretised current density are
\begin{equation}
j_{\text{src},i} = \int_\Omega J_z w_i \;\text{d}\Omega,
\end{equation}
\begin{equation}
j_{\text{pm},i}= \int_\Omega \vec{H}_{\mathrm{pm}} \cdot \left(-\parder{w_i}{y},\parder{w_i}{x}\right)\;\text{d}\Omega.
\end{equation}
\end{subequations}

The choice of the basis functions depends on the method one favours. The simplest case in the well established \gls*{fem} is the use of linear hat functions \cite{Salon_1995aa}. This paper proposes the use of the IGA framework to model the electrical machine. In this method, the basis functions are defined by \gls*{nurbs} \cite{Piegl_1997aa}.

\begin{figure}[t]
\begin{minipage}{\linewidth}
\includegraphics[width=\linewidth]{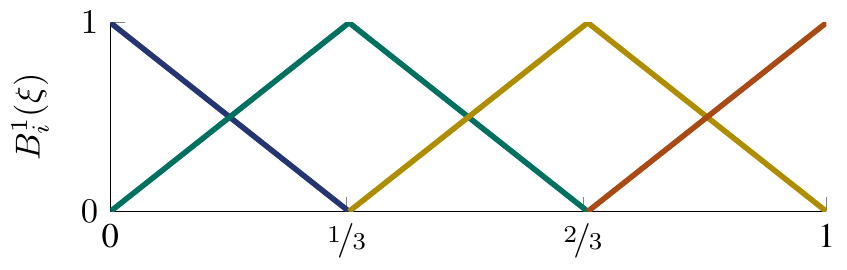}
\end{minipage}

\begin{minipage}{\linewidth}
\includegraphics[width=\linewidth]{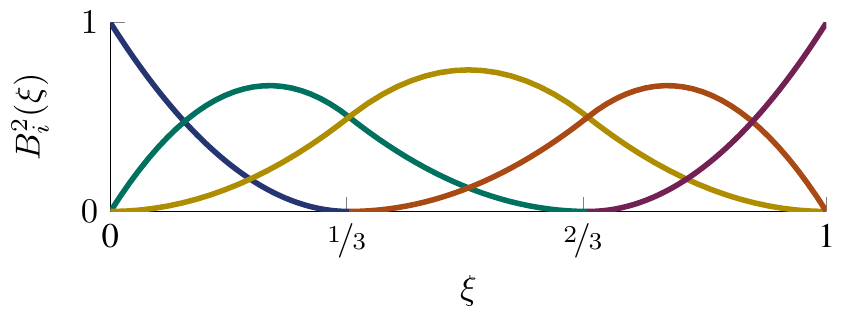}
\end{minipage}
\caption{B-spline basis functions of degree 1 and 2 on an open, uniform knot vector $\Xi = \left[0,0,0,\nicefrac{1}{3},\nicefrac{2}{3},1,1,1\right]$.}\label{fig:BSp-basis-various-deg}
\end{figure}

Let $p$ depict the degree of the basis functions and let
\begin{equation}
\Xi=\begin{bmatrix}
\xi_{1} & \dots & \xi_{n+p+1}
\end{bmatrix}
\end{equation}
be a vector that partitions $[0,1]$ into elements, where $\xi_i\in\hat{\Omega} = [0,1]$. Then, the Cox-de Boor's formula~\cite{Piegl_1997aa} defines $n$ B-spline basis functions $\lbrace B_i^p\rbrace$ with $i=1,\dots,n$. B-splines of degree $p=1,2$ are shown in Fig.~\ref{fig:BSp-basis-various-deg}. \gls*{nurbs} of degree $p$ are then constructed as
\begin{equation}\label{eq:nurbs_basis}
N_i^p = \frac{\weight_i B_i^p}{\sum_j \weight_j B_j^p},
\end{equation}
with $\omega_i$ a weighting parameter associated with the $i$-th basis function. A \gls*{nurbs} curve is obtained by the mapping
\begin{equation}\label{eq:mapping}
\mathbf{F} = \sum_{i=1}^{n} \mathbf{P}_i N_i^p,
\end{equation}
with $\mathbf{P}_i$ \emph{control points} in $\mathbb{R}^3$ and $\mathbf{F}:[0,1]\rightarrow\mathbb{R}^2$. From the curves, surfaces can be constructed by using tensor products~\cite{Piegl_1997aa}, i.e., 
\begin{equation}\label{eq:iga-basis}
\mathbf{N}_{ij}^{p_1,p_2} = N_i^{p_1} N_j^{p_2}, \quad i=1,\dots,n_1, \quad j=1,\dots,n_2.
\end{equation}
The number of one-dimensional basis functions along direction $d=1,2$ and their degrees are depicted by $n_d$ and $p_d$ respectively. The mapping $\mathbf{F}$ is now re-defined, namely $\mathbf{F} : \hat{\Omega}\to\Omega$, where $\hat{\Omega}=[0,1]^2$ is the reference square and $\Omega\in\mathbb{R}^3$ is the physical domain. Due to the \gls*{nurbs} mapping, the subdivisions constructed by the vectors $\Xi_d$ are transformed to a physical mesh on $\Omega$.

Due to the complexity of the geometry, the cross section of the machine cannot be represented by a single map $\mathbf{F}$~\cite{Cottrell_2009aa}. Multiple patches $\Omega_k$ are constructed (Fig.~\ref{fig:machine}), each one of them defined as the image of the unit square through a parametrization $\mathbf{F}_k$ of the type~\eqref{eq:mapping} in such a way that $\cup_k\Omega_k = \Omega$ and $\Omega_i\cap\Omega_j=\emptyset$ \cite{Cottrell_2009aa}. 

Using \eqref{eq:nurbs_basis} and  \eqref{eq:mapping} enables an exact parametrization of e.g. circles and arcs. This implies that the geometry of the machine can be modelled exactly, which is particular beneficial in the air gap region.
In \gls*{iga}, the same set of basis functions~\eqref{eq:iga-basis} is also used to approximate the solution of~\eqref{eq:poisson}. Hence, we can rewrite \eqref{eq:apprx} as
\begin{equation}
A_z = \sum_{j=1}^{\Ndof} u_j \mathbf{N}_j,
\end{equation} 
where we have removed the indices $p_1$, $p_2$ for simplicity. We will always assume the same polynomial degree in both directions.

This choice of basis functions guarantees a higher regularity of the solution across the elements. The continuity at the patch boundaries reverts to the \gls*{fem} case since only $C^0$ continuity is imposed (see~\cite{Cottrell_2009aa}).


\section{Iterative Stator-Rotor Coupling}
\label{sec:Coupling}
Since the rotor and the stator have a different topology, we parametrise them independently as multipatch \gls*{nurbs} entities with non-conforming patches at the air gap interface $\Gag$. The idea is to follow a classical non-overlapping domain decomposition approach based on a Dirichlet-to-Neumann map (see e.g.~\cite{Quarteroni_1999aa}).

Let us consider the circular arc $\Gag$ in the air gap and split the domain $\Omega$ such that $\overline{\Omega}_{\text{rt}} \cup \overline{\Omega}_{\text{st}} = \overline{\Omega}$ and $\overline{\Omega}_{\text{rt}} \cap \overline{\Omega}_{\text{st}} = \Gag$ (see Fig.~\ref{fig:machine}). Given $k = 1$ and an initialisation $\lambda^0$, we solve:
\begin{equation}\label{eq:DtN-Rotor}
\left\{\begin{array}{rl}
-\nabla \cdot \left(\nu \nabla \Azrt^{k+1} \right)& = J_{z},\\
\Azrt^{k+1}|_{\Gd} &=  0,\\
\Azrt^{k+1}|_{\Gl} &= -\Azrt^{k+1}|_{\Gr}\\
\Azrt^{k+1}|_{\Gag} &= \lambda^k
\end{array}\right.
\end{equation}
\begin{equation}\label{eq:DtN-Stator}
\left\{\begin{array}{rl}
-\nabla \cdot \left(\nu \nabla \Azst^{k+1} \right)& = J_{z},\\
\Azst^{k+1}|_{\Gd} &=  0,\\
\Azst^{k+1}|_{\Gl} &= -\Azst^{k+1}|_{\Gr}\\
\nu \nabla\Azst^{k+1}|_{\Gag}\cdot\vec{n}_{\text{ag}} &=
\nu \nabla\Azrt^{k+1}|_{\Gag}\cdot\vec{n}_{\text{ag}},
\end{array}\right.
\end{equation}
where $\vec{n}_{\text{st}}$ is a unit vector perpendicular to the air gap interface. The two problems are solved iteratively with the update 
\begin{equation}
\lambda^{k+1} = \alpha \Azst^{k+1} + (1-\alpha) \lambda^k,
\end{equation}
where $\alpha\in[0,1]$ is a relaxation parameter (in general, the method is not ensured to converge if $\alpha = 1$~\cite{Quarteroni_1999aa}).

As a stopping criterion for the method, the $L^2$ error between two successive iterations is required to be below a specified tolerance both in the rotor and the stator, i.e.,
\begin{align*}
\varepsilon_\text{rt} = \norm{\Azrt^{k+1} - \Azrt^k}{L^2(\Ort)} / \norm{\Azrt^{k+1}}{L^2(\Ort)} &< \texttt{tol},\\
\varepsilon_\text{st} = \norm{\Azst^{k+1} - \Azst^k}{L^2(\Ost)} / \norm{\Azst^{k+1}}{L^2(\Ost)} &< \texttt{tol}.
\end{align*}

The discretisation of problems~\eqref{eq:DtN-Rotor}-\eqref{eq:DtN-Stator} is straightforward in the \gls*{iga} framework presented above.


\section{Results}
\label{sec:Res}

\begin{figure}
\includegraphics[width=\columnwidth]{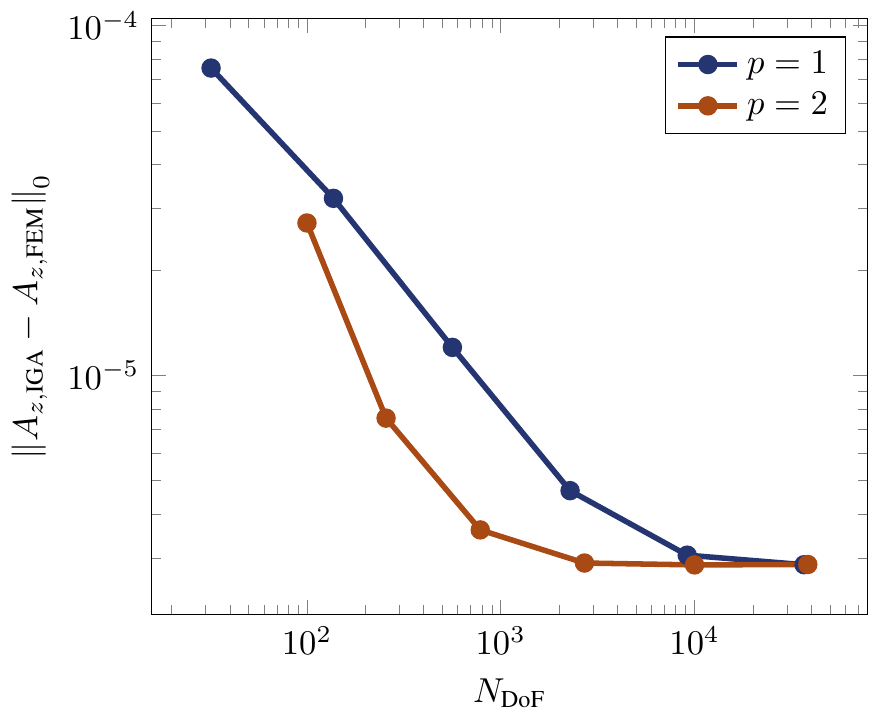}
\caption{Poisson problem for the magnetic vector potential on $\Ort$: $L^2$ error between the \gls*{iga} solution and a fine \gls*{fem} solution for increasing mesh refinement.\label{fig:iga-fem-comparison}}
\end{figure}

For testing the suitability of \gls*{iga} for machine simulation, we first consider the domain $\Ort$ and we solve a test problem with homogeneous Dirichlet boundary conditions on $\Gd\cup\Gag$ and anti-periodic boundary conditions connecting $\Gl$ and $\Gr$. The only source present in the model is the magnetisation of the permanent magnets. The \gls*{iga} solution for the magnetic vector potential is compared to the one obtained using a fine \gls*{fem} discretisation with first order elements. We solve \gls*{iga} basis functions of degree $1$ and $2$ and for an increasing mesh refinement, and we depict the difference to the \gls*{fem} solution in $L^2$ in Fig.~\ref{fig:iga-fem-comparison}. For both discretisation degrees, the curves approach a fixed value, which is expected since the \gls*{iga} method is solving the problem for the exact geometry in contrast to \gls*{fem} which introduces an additional geometrical error related to the triangulation of the geometry.

\begin{figure}
\centering
\includegraphics[width=\columnwidth]{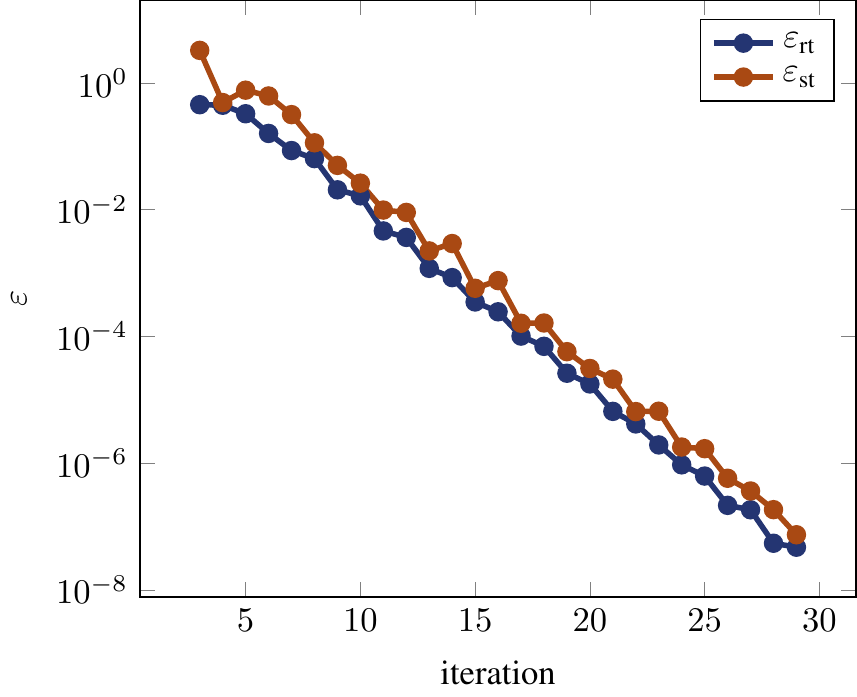}
\caption{Convergence of the Dirichlet-to-Neumann coupling for the \gls*{pmsm}. Relative difference in $L^2$ norm for the magnetic vector potential between successive iterations on $\Ort$ (in blue) and $\Ost$ (in orange).\label{fig:DtN-convergence}}
\end{figure}

Secondly, the full problem is considered, incorporating the Dirichlet-to-Neumann coupling between stator and rotor introduced in the previous section. In Fig.~\ref{fig:DtN-convergence}, the convergence of the algorithm is shown for a simulation with degree~$2$, with approximately 3200 total degrees of freedom. After 29 iterations, both solutions show an incremental error below the prescribed tolerance $\texttt{tol}=10^{-7}$.

\begin{figure}
\centering
\includegraphics[width=0.5\textwidth]{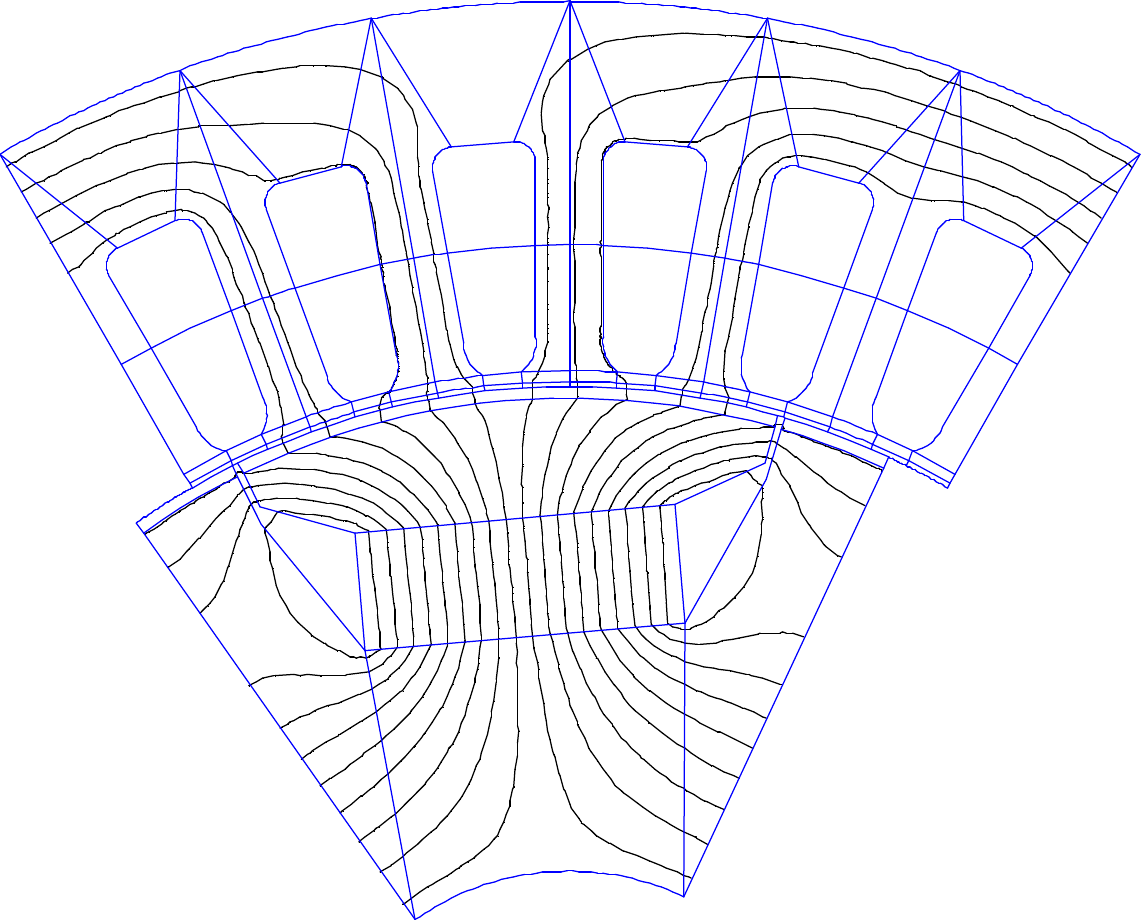}
\caption{\label{fig:fields} The distribution of the flux lines obtained by modelling with \gls*{fem}.}
\end{figure}

\begin{figure}
\centering
\includegraphics[width=\columnwidth]{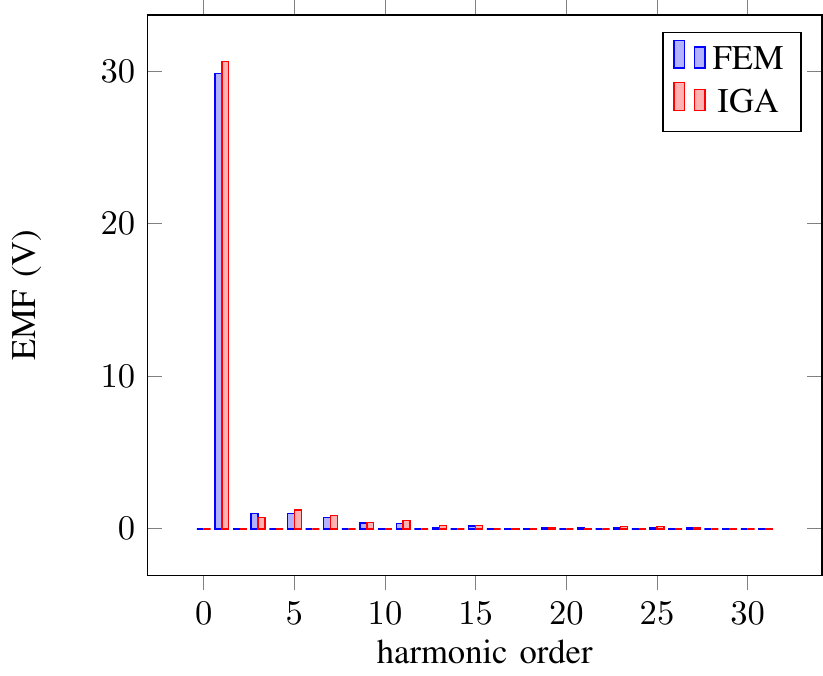}
\caption{Spectrum with the first 32 modes of the \gls*{emf} of the machine. \label{fig:spectra}}
\end{figure}

The simulation results are shown in Fig.~\ref{fig:fields}, where  the flux lines in the machine are shown. In the post-processing the spectrum of the \gls*{emf} has been calculated. In Fig.~\ref{fig:spectra}, the first $32$ modes of the spectrum of the \gls*{emf} of the machine are shown. Both methods result in similar spectra. This is also depicted in Tab.~\ref{tab:results}, where it is shown that the relative difference of the \gls*{thd} is below $6\%$.

\begin{table}
\centering
\caption{\label{tab:results}Numerical results for the \gls*{emf} and the \gls*{thd}.}
\begin{tabular}{lccc}
\hline
\hline
 & $\Ndof$& $E_1$& \gls*{thd}\\
\hline
IGA (degree 2)& 3224 & 30.6 V & 6.06$\cdot$10$^{-2}$ $\%$ \\
FEM& 59678 & 29.8 V & 5.72$\cdot$10$^{-2}$ $\%$ \\
\hline
relative difference& -& 2.5 $\%$ & 5.7 $\%$\\
\hline
\hline
\end{tabular}
\end{table}

\section{Conclusions}
Isogeometric Analysis (IGA) recently emerged as a promising alternative to the Finite Element Method (FEM) for highly accurate electromagnetic field simulation. IGA is capable of exactly resolving circles and, hence, avoids any geometric approximations errors in contrast to FEM, which is of particular importance in the air gap region. Our IGA model for a permanent magnet synchronous machine consists of two separate multipatch IGA discretisations for stator and rotor, glued together with a Dirichlet-to-Neumann map in the air gap. IGA (degree 2) attains the prescribed accuracy with a number of degrees of freedom which is substantially smaller ($\approx$ 20 times) than for lowest order FEM. Therefore, IGA is a valuable extension to an engineer's simulation toolbox, especially when fast and accurate machine simulation tasks are due.

\section*{Acknowledgment}
This work is supported by the German BMBF in the context of the SIMUROM project (grant nr. 05M2013), by the DFG (grant nr. SCHO1562/3-1) and by the 'Excellence Initiative' of the German Federal and State Governments and the Graduate School of CE at TU Darmstadt.

\ifCLASSOPTIONcaptionsoff
  \newpage
\fi

 
\end{document}